# Considerations Regarding the Flux Estimation in Induction Generator with Application at the Control of Unconventional Energetic Conversion Systems

**Iosif Szeidert, Octavian Prostean, Ioan Filip, Vasar Cristian**

Department of Automation and Applied Informatics, Faculty of Automation and Computer Sciences, "Politehnica" University from Timisoara, Av. V. Parvan, No.2, 300223, Timisoara, Romania, Phone: (0040) 256 403237, Fax: (0040) 256 403214, siosif@aut.utt.ro

*Abstract: The paper presents issues regarding the flux estimation of induction machines. The electrical machines variable's estimation represents a major problem in the actual context of modern control approaches, especially of sensorless control strategies. There are considered several implementations of induction machine's flux estimators by using the Matlab-Simulink environment and there are drawn the afferent conclusions.*

## 1 Introduction

The issue of the state estimation represents a major problem in the actual trend of modern electrical machines control. The state estimation must be solved especially in the case when those variables are not measurable or the transducers are expensive (such as torque, or flux transducers) and the price cost of a control system increases due to this fact. Therefore, appears the necessity of using of control schemes and methods without the usage of some specific transducers/sensors. This case is the one of the sensorless control methods. [1]

The estimator's implementations are based on the usage of control plant models, their aim being to estimate the value of non-measurable variables by using other measurable variables.

Basically there are two major types of estimators: [1]

- Estimators without correction (without feedback)

- Asymptotic estimators or observers (with feedback), that presents a predictive correction in order to assure a faster convergence and a better robustness of estimation at the estimated plant's parameter variations and at exogenous perturbations.

In the context of electrical machines the most commonly used estimators are:

- Flux and torque estimators – are used in the FOC (field oriented control) and the direct torque and flux vector control
- Speed and acceleration estimators based on measured position – are used in position/speed controllers, state controllers, sliding mode controllers, etc.
- Position and speed estimators based on measured stator variables (currents and voltages) – used in the of sensorless control (without position sensors/transducers)
- Perturbation estimators used in equivalent perturbation compensators.

In order to fulfill the requirements of a faster and accurate control (reduced response periods) of the induction machine, the value of the flux variable must be known. This variable's value can be estimated on the basis of voltage, current and rotation speed measurements. There are several control strategies and methods for the induction machine. In technical literature, are especially used the control based on stator flux and the one based on the rotor flux. [2] [3].

The paper describes some estimator structures implemented in the Matlab-Simulink simulation environment. The paper studies those estimator structures having in view the possibility of their implementation in unconventional energetic conversion systems, especially of WECS (wind energy conversion systems). There are studied three induction machine's flux estimators: based on voltage model, based on current model and based on the stator current estimation.

## 2  Induction Machine's Flux Estimators

### Flux Estimator based on Voltage-model

This type of estimator (flux estimator based on voltage model) can be synthesized by using the voltage-based model, as depicted in relation (1):

$$\hat{\Psi}_s = \int (u_s - R_{se} i_s) dt \tag{1}$$

Where: ˆ - represents the estimated value,

    e – represent the estimator parameter,

$\hat{\Psi}_s$ - the stator flux estimated value,

$u_S$ – stator voltage,

$R_{SE}$ – the stator resistance estimator's parameter,

$i_S$ – stator current.

There is considered only the real part of the equation and respectively a zero value rotor current (without load torque), the obtained results are valid also in the case of the imaginary axis (with load torque).

The estimator is defined by the relation (2), for the x-axis:

$$\hat{\Psi}_{sx} = \int (u_{sx} - R_{se} i_{sx}) dt \qquad (2)$$

The relation (2) implementation in Matlab-Simulink is represented in Figure 1.

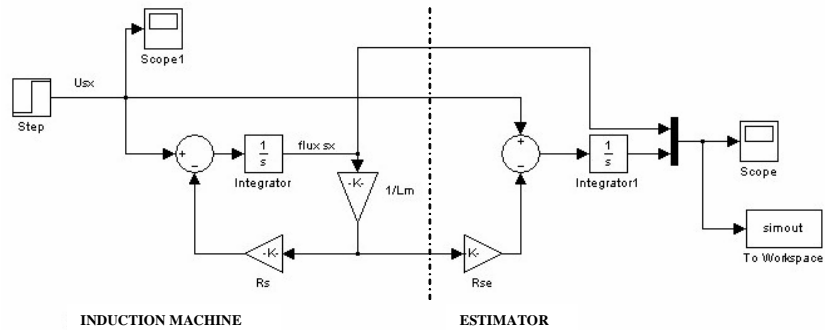

Figure 1

Induction machine's flux estimator based on the voltage model (Matlab-Simulink implementation)

The input variables in the estimator are the machine's current and voltage. In the case that the value of the estimator's parameter stator resistance $R_{SE}$ is identical to the electrical machine stator resistance $R_S$ and the current and voltage are measured without errors (due to noise or offset) then the estimator can be used when the rotor current is not zero. This fact can be noticed in relation (1) where the rotor current doesn't occur.

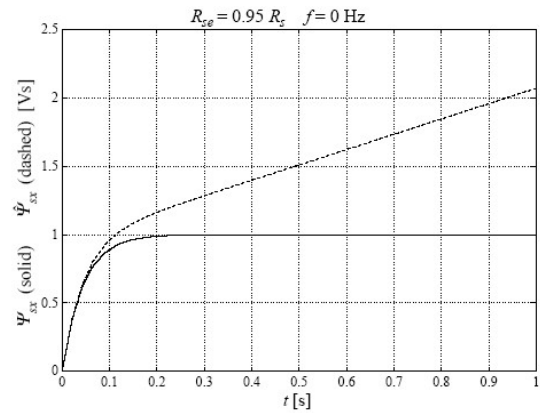

Figure 2
Simulation results (estimator's error)

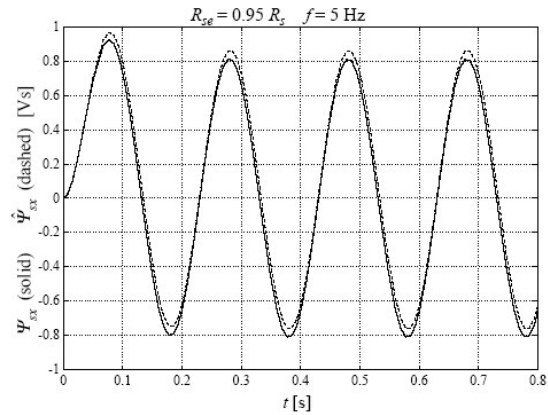

Figure 3
Simulation results (estimator's error at f=5 Hz)

However, in the case of a slightly different value of the $R_{SE}$ parameter this would lead to errors in the estimator performance especially in the case of low rotation speeds. In Figure 2 is represented the estimator's error in the case when the resistance value is different with a 5% regarding the real value of the resistance (the response at a step signal – the $u_{sx}$ voltage value). It can be noticed, that initially the estimate follows the flux evolution, but further it has a deviation to infinite or to the integrator saturation.

In Figure (3) there can be noticed that the estimate follows correctly the flux variable, even in the case of a 5 Hz frequency (the resistance error being also 5%). There can be concluded that the influence of the resistance parameter error reduces with the frequency increase. Also, the offset in the current measurement conducts to errors in the flux estimation at any frequency value. Taking in consideration that this estimation method presents difficulties in implementations (especially al low rotation speed), practically it is not used in applications.

**Flux Estimator based on Current-model**

In order to improve to performances of the previous estimator there is used the relation (3).

$$\Psi_s = L_M (i_s + i_r) \tag{3}$$

Where: $L_M$ – magnetization inductance,

$i_s$ –stator current,

$i_r$ – rotor current,

$\Psi_s$ –stator flux

There is considered the rotor current to be zero, and from the relation (3) results the following relation (4):

$$\hat{\Psi}_{sx} = L_{me} i_{sx} \tag{4}$$

Where: $L_{me}$ – the magnetization inductance estimator's parameter.

Therefore, only the stator current represents the input of the estimator structure. In figure 4, there is represented the Matlab-Simulink implementation of induction machine's flux estimator based on current-model.

One of the main disadvantages of this estimator implementation based on the current-model is the fact that a correct estimation of flux requires a very precise value of the magnetization inductance $L_M$. Another problem is the fact that due to the magnetic saturation phenomenon that occurs in the electrical machines, this parameter ($L_M$) is not constant. This parameter varies, fact that leads to a difficult obtaining of the $L_{me}$ parameter. In Figure 5 there are represented the simulation results – the evolution of the flux estimate in the condition of a 5% error in the magnetizing inductance value. Another disadvantage of this estimator type is the fact that the errors are significant in the case that the rotor current is not zero. In order to compensate the rotor current effect, it is necessary to measure the rotation speed because the rotor current usually cannot be measured.

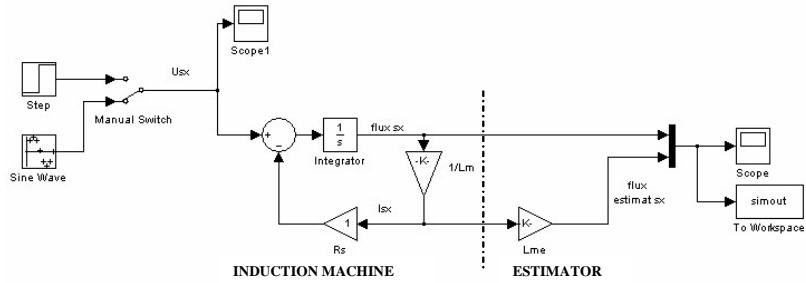

Figure 4

Induction machine's flux estimator based on the current model (Matlab-Simulink implementation)

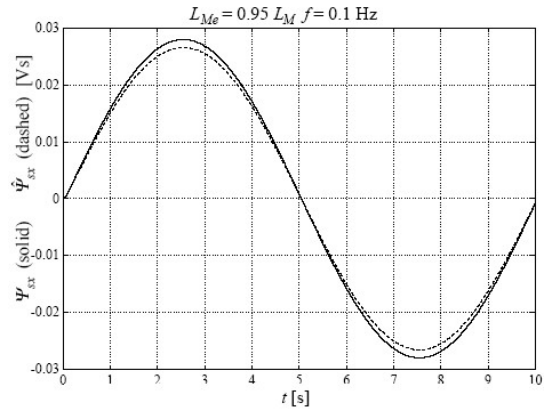

Figure 5

Simulation results (estimator's error)

The previous estimator, based on the voltage-model presents the advantage of being independent to the rotation speed and to the rotor current.

From relation (5) and relation (3) and (4) there is obtained the relation (6).

$$\Psi_r = \Psi_s + L_L i_r \tag{5}$$

Where: $L_L$ – is leakage inductance.

$$\Psi_s = (L_L i_s + \Psi_r) \frac{L_M}{L_L + L_M} \tag{6}$$

In order to calculate the rotor flux, from relation (5), (4), (3) and (7) there is obtained the relation (8).

$$\frac{d\Psi_r}{dt} = jz_p\omega\Psi_r - R_r i_r \tag{7}$$

where: $z_p$ – number of poles pairs,

$\omega$ – rotor rotation speed,

$R_r$ – rotor resistance.

$$\frac{d\Psi_r}{dt} = i_s \frac{R_r L_M}{L_L + L_M} - \Psi_r \left(\frac{R_r}{L_L + L_M} - jz_p\omega\right) \tag{8}$$

There results the estimator defined by the relations (9) and (10):

$$\hat{\Psi}_s = (L_{Le} i_s + \hat{\Psi}_r) \frac{L_{Me}}{L_{Le} + L_{Me}} \tag{9}$$

$$\frac{d\hat{\Psi}_r}{dt} = i_s \frac{R_{re} L_{Me}}{L_{Le} + L_{Me}} - \hat{\Psi}_r \left(\frac{R_{re}}{L_{Le} + L_{Me}} - jz_p\omega\right) \tag{10}$$

This estimator has as input the rotor rotation speed and the stator current, as it can be noticed in Figure 6.

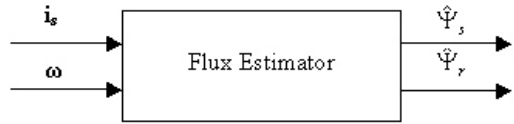

Figure 6
Flux estimator based on current-model

The flux estimator based on current-model presents good performances in the case of low frequencies, but is sensible to the parameter error at high frequencies.

**Flux Estimator based on the Stator Current Estimation**

In order to solve the problem of the estimator's based on voltage-model error, there is used the estimate of the stator current instead of the measured value of the stator current variable.

$$\frac{d\Psi_s}{dt} = u_s - R_s i_s \tag{11}$$

From relation (11) and (3), considering the rotor current to be zero, there results the estimator described by relations (12) and (13), represented in Figure 7.

$$\hat{\Psi}_{sx} = \int (u_{sx} - R_{se}\hat{i}_{sx})dt \tag{12}$$

$$\hat{i}_{sx} = \frac{\hat{\Psi}_{sx}}{L_{Me}} \tag{13}$$

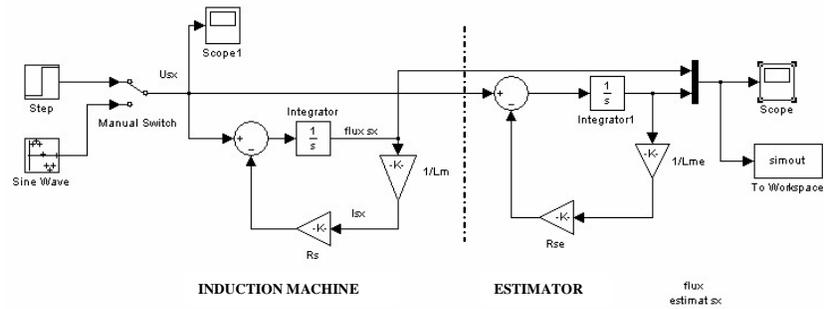

Figure 7

Induction machine's flux estimator based on stator current estimation. (Matlab-Simulink implementation)

This flux estimator is principally a simulation model in open loop, because there are no feedbacks through some measurements. In Figure 8, there can be noticed the reduced value of the estimator's error. However, there is present a steady state regime error in the flux estimate due to the error in the resistance or inductances errors.

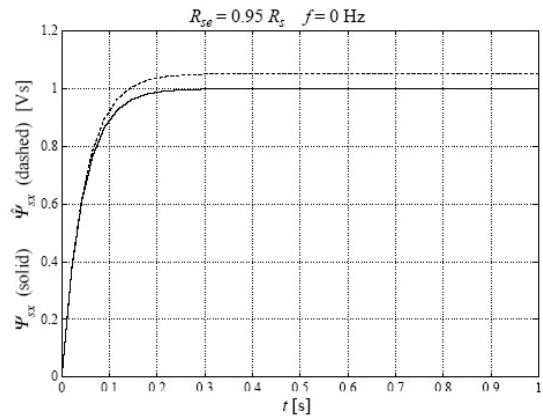

Figure 8

Simulation results (estimator's error)

**Conclusions**

There can be concluded that the solution of using the induction machine's magnetic flux estimation is imposed because this variable must be known in the context of the control structure, especially in the case of FOC structures and due to the fact that this variable cannot be directly measured.

The three estimator structures presented in the paper present both advantages and limitations that must be considered and studied for each type of application. The problem that must be solved is the design of a flux estimator structure that presents good performances in the case of entire frequency range spectrum. The first presented estimator based on voltage-model presents a better response in the case of higher frequency range; meanwhile the flux estimator based on current-model presents a better response in the range of low frequencies.

In [1] [4], there is described an efficient modality of a combination of those two estimators in order to obtain a estimator with good performances both in case of low frequency range and of high frequency range. The idea that leads to a such estimator is to use a low frequency pass filter to select the estimator based on the current-model at low frequency range and of course, a high frequency pass filter to select the estimator based on voltage-model at high frequency range.

Another approach modality that can be used is the observer theory, such as the Kalman filtering that can be applied also in the case of the induction generator. [5] [6]

The above exposed flux estimators present a great significance in the actual trend of the sensorless control structures of the modern windmills based on induction generators. In order to study the windmills control structure performances there is mandatory a detailed analysis of the dynamic behavior of all main components of the wind energy conversion line: wind turbine, electrical generator, converter, grid and other additional elements).